\documentclass[nofootinbib,reprint,amsmath,amssymb,showkeys, showpacs,aps]{revtex4-1}
\usepackage{graphicx}
\usepackage[T1]{fontenc}
\usepackage{dcolumn}
\usepackage{xcolor,colortbl}
\usepackage{multirow}
\usepackage{stackengine}
\usepackage{tikz-cd}
\usepackage{bm}
\usepackage{mathtools}
\usepackage{tikz}

\usepackage{hyperref}
\hypersetup{colorlinks, linkcolor={black},citecolor={blue},urlcolor={blue}}

\usepackage{accents}
\usepackage{scalerel}
\usepackage{bbm}
\usepackage{tabularx,siunitx}
\usepackage{wasysym}
\usepackage{ mathrsfs }
\usepackage{bbold}

\begin{document}

\preprint{APS/123-QED}

\title[The Equivalence Principle as a Noether Symmetry]{The Equivalence Principle as a Noether Symmetry}

\author{Salvatore Capozziello$^{1,2,3}$}\email{capozziello@unina.it}
\author{Carmen Ferrara$^{2,3}$}\email{carmen.ferrara-ssm@unina.it}
\affiliation{$^1$ Dipartimento di Fisica "E. Pancini", 
Universit\`a degli Studi di Napoli ``Federico II'', Via Cinthia Edificio 6, 80126 Napoli, Italy\\
$^2$ Scuola Superiore Meridionale, Largo San Marcellino 10, I-80138 Napoli, Italy,\\
$^3$ Istituto Nazionale di Fisica Nucleare, Sezione di Napoli, Complesso Universitario di Monte S. Angelo, Via Cinthia Edificio 6, I-80126 Napoli, Italy}

\date{\today}

\begin{abstract}
The Equivalence Principle is considered in the framework of metric-affine gravity. We show that it naturally emerges as a Noether symmetry starting from a general non-metric theory. In particular, we discuss the Einstein Equivalence Principle and the Strong Equivalence Principle showing their relations with the non-metricity tensor. Possible violations are also discussed pointing out the role of non-metricity in this debate.
\end{abstract}

\maketitle

\section{Introduction}

Gravity is one of the  interactions of Nature but its formulation, at fundamental level, is still  matter of debate. In 1915, Einstein published the theory of  General Relativity (GR) as a geometric  formulation of gravitational interaction apt to  account for the motion of astrophysical bodies and  the dynamics of the Universe. The foundation of the theory is based upon some principles \cite{Romano2019}: the General Covariance, the Equivalence Principle  (EP), and  the Principle of Causality. 
According to them, the space-time structure is determined by two  fundamental  fields, a Lorentzian metric $g$ (fixing the space-time causal structure) and a linear connection $\Gamma$ (defining the free-fall of the bodies). $\Gamma$ and $g$ can be a-priori independent, but, in GR, $\Gamma$ has to be the Levi-Civita connection constructed upon the metric tensor $g$ and then defining  the locally inertial observers in agreement with EP).

The gravitational interaction is geometrically encoded in the curvature of a four-dimensional space-time manifold $(M,g)$. Curvature is locally determined by the distribution of  sources, which are expressed by the four-dimensional generalization of the ”matter stress-energy tensor”, e.g., a rank-two (symmetric) tensor $T^{\mu\nu}$.
The affine linear connection and parallel transport enclose their physical interpretation in EP, about which there are some different versions. Here we mainly focus on two of them: (1) the Einstein Equivalence Principle (EEP), postulating the equivalence between gravitational and inertial effects; (2) the Strong Equivalence Principle (SEP), according to which GR is locally special relativistic, i.e., locally there is no evidence of gravity. Physically, the EEP means that gravity is the  space-time geometry, so that   dynamics is governed by the Local Lorentz invariance (LLI), the Local Position Invariance (LPI), and the Weak Equivalence Principle (WEP). According to the so-called {\it Schiff conjecture}, the validity of EEP depends on the simultaneous validity of the three aforementioned conditions \cite{Schiff60}. On the other hand, the SEP states that: \emph{all tests of fundamental physics, including gravitational physics, are not affected, locally, by the presence of a gravitational field.} 

Specifically, the SEP guarantees the local validity of Special Relativity (SR) in GR, and, since it does not follow from the Einstein field equations, it provides the geometric interpretation of the metric $g_{\mu\nu}$ beyond the gravitational dynamics. The EEP allows to unify the Newtonian concepts of gravity and inertia. In other words, it is the bridge between GR and Newtonian gravity, as well as the SEP is the bridge between GR and  Special Relativity (SR). Following Refs. \cite{Pauli21, Brown07}, the SEP characterizes the internal structure of GR itself.

However, the SEP results to be valid also for the Teleparallel Equivalent of GR (TEGR) and the Symmetric Teleparellel Equivalent of GR (STEGR). These theories belong to the more general class of Metric Affine Theories, in which the Riemaniann geometry can be extended. The gravitational field is expressed with respect to other two dynamical variables, the torsion and the non-metricity tensors, since the connection represents another geometrical field with its own dynamics, physical implications, and independent of the metric. In this framework,  the Poincar\'e group can be enlarged to space-time symmetries giving rise to  the so-called Metric Affine Gravity (MAG), where the spin density, the dilation, and the shear currents are properties of  matter associated to the connection dynamics together with the energy-momentum tensor
\cite{Hehl76, Hehl376, Obukhov93}. 

Thus, in TEGR and STEGR, the SEP is no more at the foundation of the theory, as in GR, but it is a  consequence of the Poincaré gauge.  These three theories form the so-called {\it Geometric Trinity of Gravity} because, despite of the differences in the representations, field equations give the same dynamics \cite{Heisenberg19, Ferrara22, Jimenez18}.
In general,  metric theories of gravity, like  GR or the Jordan-Brans-Dicke one \cite{Brans61, Kajari10},  satisfy both EEP and SEP. In these theories, the metric and  energy-momentum tensors  can involve also other fields  associated with gravity, such as scalar or vector fields. On the other hand, non-metric theories can be defined with  variable non-gravitational couplings, associated to dynamical fields coupled to matter. Thus, in non-metric theories, the gravitational coupling to other non-gravitational fields (i.e.,  particle fields of the Standard Model) is not universal, while in metric theories, the coupling to the gravitational field is universal \cite{Will93}. As a consequence, experimental tests of the EEP are often performed considering, a priori,  the universal coupling of gravity, where different non-gravitational fields and particles are used to study the metric aspects of space-time, i.e., the mutual attraction between  bodies or the detection of  gravitational waves. Conversely, in the case of SEP, we have that the observers of an experiment will locally obtain the same results, independently of the background field. In fact, throughout the universal coupling to the gravitational field,  test particles  behave in the same way both in a gravitational field and in vacuum, independently of their properties. 

In this debate, it is worth noticing the paper Ref.\cite{Jabbari}, where a generic metric-affine theory of gravity is considered. Authors
argue that EEP requires metric and metric-affine
(Palatini) formulations to be equivalent. Then, they prove that only for Lovelock theories, metric and Palatini formulations are equivalent.  In this sense, only Lovelock theories satisfy EEP.

In particular,  a universal coupling for the gravitational field has been assumed in several  efforts  to unify GR and Quantum Mechanics (QM) in view  of Quantum  Gravity (QG).    Moreover, the validity of the Superposition Principle, required in QM as well as in QG,  allowed several studies on gravity-matter entanglement \cite{Bernard98, Oniga16, Paunkovic18, Bose17},  indefinite causal orders \cite{Oreshkov19, Vilasini19}, quantum reference frames \cite{Giacomini17, Giacomini20}, and deformations of Lorentz symmetry \cite{Camelia12, Camelia20}.

Present-day tests of  EEP are sensitive only to violations that alter the spectra of internal energies. Such violations can be expressed in terms of differences between diagonal elements of  operators. In fact,  it is assumed that internal energy operators commute, so that EEP violations regard only their eigenvalues \cite{Zych18}. This led to several formulations of EEP in the context of QG, and, in particular, of the WEP, which  could be violated for some specific  gravitational field superpositions \cite{Pipa19,Das23}. 

Some space missions have been proposed with the goal to validate EEP, also in the QM contest: for example, the  "MICROSCOPE mission" \cite{MICROSCOPE19}, the "Satellite Test of the Equivalence
Principle" (STEP) \cite{Sumner07},  the ”Gamma Ray Astronomy International Laboratory for QUantum Exploration of Space–Time” (GrailQuest) \cite{Burderi21},  the ”Space Atomic Gravity Explorer” (SAGE) \cite{Tino20}, the ”Space-Time Explorer and QUantum Equivalence Space Test” (STE-QUEST) \cite{Altschul14}.
In most of them, however, the gravitational coupling is assumed universal.

In this paper,  we want to  demonstrate that EEP can be derived from the Noether theorems and, as consequence, is a Noether symmetry. On the contrary, if the Noether symmetry does not hold, also the EEP can be violated. 

The layout of the paper is the following.  In Sec \ref{sec:Metric Affine}, we describe the MAG framework. The role of  linear affine connection is highlighted in Sec \ref{sec:parallel}. The two Noether theorems   and their geometrical interpretation are reported in Sec. \ref{sec:Noether theorems}. Finally,  in Sec. \ref{sec:EP and Noether Symmetry}, we show that the EP is a Noether symmetry. Conclusions are drawn in Sec. \ref{Conc}.

\textit{Notation:} Quantities with tilde are built up in the Levi-Civita connection (i.e. $\tilde{A}_\mu$). We use the Lorentz metric with signature, $(-, +, +, +)$. It is denoted with $g_{\mu\nu}$ or $g$. Greek indices represent the space-time, with values $ 0, 1, 2, 3$. Latin indices indicate general coordinates (internal and external) not necessarily space-time ones. The flat metric is indicated by $ \eta_{\alpha\beta} = \text{diag} (-1, 1, 1, 1)$.

\section{Metric-Affine Theories of Gravity}
\label{sec:Metric Affine}

GR is based on the Riemaniann geometry,  where the metric tensor $g_{\mu\nu}$ plays the role of fundamental field. For this reason, the associated linear connection is metric-compatible, i.e., length and angle measurements are integrable and symmetric. The symmetry of the Levi-Civita connection in the lower indices leads to the closure of infinitesimally small parallelograms during the parallel transport \cite{Eisenhart97}. Therefore, a natural way to modify and/or extend the geometry is to relax the Riemannian constraints. In other words, by starting with a general linear affine connection  admitting both torsion and non-metricity, it is possible to enlarge the geometric framework, considering  non-Riemannian geometries \cite{Eisenhart12}. The resulting gravity theory is the MAG, that is a class of theories where metric and affine connection are both fundamental fields \cite{Saridakis21}. 
In this framework, the manifold is described by the following invariant quantities: curvature, torsion, and non-metricity. These quantities are described by the metric $g$ and affine connection $\Gamma$, having thus the structure $(M, g, \Gamma)$. The connection coefficients can be uniquely decomposed, with respect to a given metric, as: 
\begin{equation}
    \label{affine connection}
    \Gamma^{\lambda}_{\mu\nu} = \tilde{\Gamma}^{\lambda}_{\mu\nu} + K^{\lambda}_{\mu\nu} + L^{\lambda}_{\mu\nu},
\end{equation}
where $\tilde{\Gamma}^{\lambda}_{\mu\nu}$ is the Levi-Civita connection, $K^{\lambda}_{\mu\nu}$ is the contortion tensor, and $L^{\lambda}_{\mu\nu}$ is the disformation tensor, whose explicit expressions are:
\begin{subequations}
\begin{align}
\tilde{\Gamma}^{\lambda}_{\mu\nu}&:=\frac{1}{2}g^{\rho\lambda}(\partial_\mu g_{\rho\nu}+\partial_\nu g_{\mu\rho}-\partial_\lambda g_{\mu\nu}),\label{eq:LC}\\
K^\lambda_{\ \mu\nu}&:=\frac{1}{2}(T_\mu^{\ \lambda}{}_\nu+T_\nu^{\ \lambda}{}_\mu-T^\lambda_{\ \mu\nu}),\label{eq:contortion}\\
L^\lambda_{\ \mu\nu}&:=\frac{1}{2}(Q^\lambda_{\ \mu\nu}-Q_\mu^{\ \lambda}{}_\nu-Q_\nu^{\ \lambda}{}_{\mu})\label{eq:deformation}.
\end{align}
\end{subequations}
Therefore, the three dynamical (geometric) variables, built on $\Gamma$, have the following explicit expressions in terms of metric and linear connection:
\begin{subequations}
\begin{align}
R^{\alpha}_{\beta\mu\nu}&:= \partial_{\mu}\Gamma^{\alpha}_{\beta\nu}-\partial_{\nu}\Gamma^{\alpha}_{\beta\nu} + \Gamma^{\alpha}_{\sigma\mu}\Gamma^{\sigma}_{\beta\nu}-\Gamma^{\alpha}_{\sigma\nu}\Gamma^{\sigma}_{\beta\mu},\label{eq:curvature}\\
T^\alpha_{\ \mu\nu}&:= \Gamma^\alpha_{\ \mu\nu}-\Gamma^{\alpha}_{\ \nu\mu},\label{eq:torsion}\\
Q_{\alpha\mu\nu}&:= \partial_{\alpha}g_{\mu\nu} - \Gamma^{\sigma}_{\alpha\mu}g_{\alpha\nu}-\Gamma^{\sigma}_{\alpha\nu}g_{\sigma\mu}\label{eq:nonmetricity}.
\end{align}
\end{subequations}
They affect the parallel transport of a vector defined by the linear affine connection. During the parallel transport along a closed curve on a non-flat background, curvature causes a non-null angle when the vector comes back in its initial position; torsion expresses the antisymmetric behaviour of two vectors when they and their direction are switched, thus it represents how much the formed parallelogram is not-closed; non-metricity causes the change of the norms and the lengths of vectors under the parallel transport \cite{Bahamonde23, Heisenberg23}.

In MAG, some of the three dynamical variables can be trivial and then it is possible to classify MAGs in the following  theories:
\begin{itemize}
    \item Metric theories defined in  terms of Riemannian geometry given by the couple $(M,g)$. They are the dynamical  arena of GR, $f(R)$ gravity, and of other theories like Brans-Dicke or scalar-tensor theories. Here, dynamics is given  in terms of curvature, constructed only by the metric \cite{CapBook};
    \item Metric-affine theories {\it à la Palatini} where $g$ and $\Gamma$ are disentangled and geometry is given by the triplet $(M,g,\Gamma)$. Here, dynamics results enlarged because both $g$ and $\Gamma$ are fundamental fields. In this case, while GR is exactly restored, other theories, like $f(R)$ or scalar-tensor ones result different with respect to the pure metric analogues \cite{Olmo:2011uz};
    \item Teleparallel geometries, where the parallel transport of vectors is independent of the path, the fundamental fields are the {\it vierbeins} and affinities are given by the Weitzenb\"ock connection \cite{Aldrovandi13}.
\end{itemize}
Among teleparallel geometries, it is worth mentioning  TEGR and STEGR, since, together with GR, they constitute the so-called {\it Geometric Trinity of Gravity} \cite{Heisenberg19}, because it can be proved that they are dynamically equivalent at  Lagrangian, field equations and solutions levels \cite{Ferrara22}.

On the other hand,  symmetric teleparallel theories have been  introduced in terms of differential forms \cite{Nester98, Hehl94}. They represent a subclass of  teleparallel theories, where the constraint related to the metric postulate  is relaxed. In fact,  metric and  linear affine connection coefficients transform under a general coordinate map $\xi^{\lambda}\to x^{\lambda}$ as:
\begin{align}
   g_{\mu\nu}(x^\lambda) &= \frac{\partial \xi^{\alpha}}{\partial x^{\mu}} \frac{\partial \xi^\beta}{\partial x^{\nu}} g_{\alpha\beta}(x^\lambda),\\
   \Gamma^{\rho}_{\mu\nu}(x^\lambda) &= \frac{\partial x^{\rho}}{\partial \xi^{\gamma}} \frac{\partial \xi^{\alpha}}{\partial x^{\mu}} \frac{\partial \xi^\beta}{\partial x^{\nu}} \Gamma^{\gamma}_{\alpha \beta}(\xi^{\lambda}) + \frac{\partial^{2} \xi^{\alpha}}{\partial x^{\mu \partial x^{\nu}}} \frac{\partial x^{\rho}}{\partial \xi^{\alpha}}.\label{eq:transf coordinate CG}
\end{align}
Therefore, it emerges that the curvature scalar associated to the general curvature tensor  (\ref{eq:curvature}) is:
\begin{equation}\label{eq:curvscalar}
    R = \tilde{R} - Q + B
\end{equation}
where the non-metricity scalar is defined as:
\begin{equation}
    Q = \frac{1}{4} \big(Q_{\alpha}Q^{\alpha} - Q_{\alpha\beta\gamma}Q^{\alpha\beta\gamma} \big) + \frac{1}{2}\big(Q_{\alpha\beta\gamma}Q^{\beta\alpha\gamma}-Q_{\alpha}\bar{Q}^{\alpha}\big)
\end{equation}
with $Q_{\alpha}:=Q_{\alpha\lambda}{}^{\lambda}$ and $\bar{Q}_{\alpha}:= Q^{\lambda}{}_{\lambda\alpha}$.  \emph{B}  is a boundary term referred to the non-metricity scalar Q \cite{jimenez19, Bahamonde22, Ferrara22, Jimenez18}. 

In the symmetric teleparallel geometry, it is possible to choose a coordinate system where the connection vanishes globally on the manifold. This particular system of coordinate is the so-called \emph{coincident gauge}, since the connection is trivial. In this case,  covariant and partial derivatives coincide.  The physical meaning of the coincident gauge is that the origin of space-time, whose coordinate are $x^{\mu}$, coincide with the origin of  tangent space, whose coordinate are $\xi^{\alpha}$, namely we apply a Poincaré (linear affine) transformation. When the coordinate system, corresponding to the coincident gauge is chosen, then the transformation (\ref{eq:transf coordinate CG}) of the connection into any other coordinate system $x^{\alpha}$ reduces to:
\begin{equation}
    \Gamma^{\rho}_{\mu\nu}(x^\lambda)=\Big(\frac{\partial \xi^{\alpha}}{\partial x^{\rho}} \Big)^{-1} \partial_{\mu} \Big( \frac{\partial \xi^{\alpha}}{\partial x^{\nu}} \Big),
\end{equation}
since the following affine (gauge) roto-translational transformation of coordinates has been performed:
\begin{equation}
    \xi^\alpha := M^{\alpha}_{\beta}x^\beta + \xi^\alpha_0.
\end{equation}
In this gauge, the field equations are equivalent to GR because:
\begin{equation}
    \tilde{R}= Q - \tilde{B}
\end{equation}
and the boundary term $B$ assumes the form:
\begin{equation}
    \tilde{B} = \tilde{\nabla}_{\alpha}(Q^{\alpha} - \bar{Q}^{\alpha})\,.
\end{equation}
See \cite{DeFalco23} for details.

\section{Connection and Parallel Transport} \label{sec:parallel}

Gravitation, in the GR picture,  is a manifestation of space-time curvature, and that curvature shows up in the deviation of a geodesic from a nearby one (i.e., relative acceleration of test particles) \cite{Gravitation2000}. In this scenario,  the EEP has the role to connect flat and curved space-times. In fact, according to the Principle of General Covariance \cite{Weinberg72}, an equation valid in SR holds in presence of a gravitational field if it is general covariant. Thus, SR laws can be locally recovered when the metric of a curved space-time is replaced by the Minkowski one, that is  $g_{\mu\nu} \rightarrow \eta_{\mu\nu}$,  and the equations do not change when we perform a general coordinate transformation $x \rightarrow x^{'}$.
In order to make an equation generally covariant, we have to introduce an extra structure called {\it connection}, which specifies how tensors are transported along a curve. In general, it represents the inertial properties of the coordinate system under consideration. 

The Principle of General Covariance can be thought as an active version of EEP: given an equation valid in presence of gravitational interaction, the corresponding SR equation is locally recovered (at a point or along a trajectory). At the same time, SR equations must be recovered in a locally inertial frame (passive version) \cite{Sciama64}. 

A general linear affine connection $\nabla$ on a manifold $M$ is a map, which assigns to every pair smooth vector fields $X,Y$ on $M$ another smooth vector field $\nabla_X Y$. In a local chart  $(U, \phi)$ on $M$, with coordinate $x=\phi(p)$, the two vector fields  read as $X=X^i e_i$ and $Y=Y^i e_i$, where $\{e_i\} = \partial/ \partial x^i$ is the coordinate basis of the tangent space at any point of the open set $U$. The coordinate representation of the map $\nabla$ is:
\begin{align}\label{eq:coeffconnessione}
    &\nabla_X Y= X^i \Big(\frac{\partial Y^k}{\partial x^i} + Y^j \Gamma^k_{ij}\Big) e_k\equiv X^k Y^i_{;k} e_i, 
\end{align}
where $\nabla_{e_k}(e_j) = \Gamma^i_{kj} e_i$ are the \emph{connection coefficients}. In particular, they have not to satisfy a priori the symmetry condition $\Gamma^{k}_{ij}=\Gamma^{k}_{ji}$.
The covariant derivative represents the  generalization of  directional derivative of functions to tensors. 

Given a curve $\gamma$ on a manifold $M$, let us consider a chart $(U,x^i)$ on $M$ and the parametric equations of $\gamma$, $\gamma^i(\tau)$, with $\tau$ the affine parameter. We define the tangent vector $\dot{\gamma}$ to the curve $\gamma$ in the natural basis ${e_i}$ as follows:
\begin{equation}
    \dot{\gamma} \equiv \frac{d \gamma}{d \tau} e_i = \frac{d \gamma}{d \tau} \partial_i.
\end{equation}
Let $Y$ be a vector field defined on an open neighborhood of $M$. $Y$ is said to be parallel transported along $\gamma$ if:
\begin{equation}
    \nabla_{\dot{\gamma}} Y = \frac{d Y}{d \tau} = 0,
\end{equation}
or, in components:
\begin{equation}
    \nabla_{\dot{\gamma}} Y^k =\frac{d Y^k}{d \tau} + \Gamma^k_{ij} \frac{d \gamma^i}{d \tau}  Y^j=0.
\end{equation}
Notice that $Y(\gamma(\tau))$, obtained by a parallel transport of $Y(\gamma(\tau_0))$ along $\gamma$, depends on $\gamma$, with initial condition $Y^i(\gamma(\tau_0))=Y^i_0$ \cite{Romano2019}.
If the tangent vector $\dot{\gamma}(\tau)$  is parallel transported along $\gamma(\tau)$, namely if:
\begin{equation}\label{eq:partra}
    \nabla_{\dot{\gamma}} \dot{\gamma} = \frac{d^2 x^k}{d\tau^2} + \Gamma^k_{ij}\frac{dx^i}{d\tau} \frac{d x^j}{d\tau}=0,
\end{equation}
with $x^i$ coordinates of $\gamma(\tau)$ in the chart $(U, \phi)$, then the curve $\gamma$ is said to be autoparallel. When the affine connection is the Levi-Civita one, i.e. in GR, the autoparallel curve is said geodesic. In teleparallel gravity, they give rise to two different structures, because  autoparallels are related to the affine connection, whereas the geodesics are related to the metric, since they measure the minimal lengths between two or more points. 

These considerations can be linked together under the standard of the two  Noether theorems. We will show that the same  EP is a Noether symmetry under given conditions.

\section{The Noether Theorems}
\label{sec:Noether theorems}

In 1918, Noether discovered a relation between (continuous families of) symmetries of Lagrangian systems and their first integrals \cite{Noether18}. Conserved quantities allow to reduce and integrate dynamics.

The first Noether theorem states: {\it If a dynamical system is defined on a manifold $M$ of dimension $m={\rm dim}(M)$, then there exists a $(m-1)-$form on $M$, called  Noether current. This quantity results to be closed along solutions, implying a continuity equation. Therefore  conserved quantities are defined as the integrals of such currents on a $(m-1)-$volume in $M$. }

The continuity equation holding for the Noether current relates the changes of conserved quantities to the flows at the boundary of the region and some residual at singularities. In GR, Noether currents are not only closed forms, but even exact forms along solutions. This introduces a superpotential $S$ for each Noether current and conserved quantities are obtained by surface integrals of $S$ \cite{Fatibene10, bajardi_capozziello_2022}.

Given a classical $n$-degree-of-freedom dynamical system of particles, an infinitesimal point transformation maps “points” in configuration space and time into infinitesimal neighboring “points” \cite{Struckmeier02}. Thus, the first Noether theorem relates the conserved quantities of an $n$-degree-of-freedom Lagrangian system $L(q, \dot{q}, t)$ to infinitesimal point transformations that leave the Lagrange action $Ldt$ invariant. It formally states that: 

\emph{To every differentiable symmetry, generated by local actions, there corresponds a conserved current}, called \emph{Noether current}.

If we consider a general Lagrangian density $\mathcal{L}$, depending on coordinates $x^a$ and fields $\phi^i$, we get \cite{bajardi_capozziello_2022}:
\begin{subequations}
\begin{align}
   \mathcal{L}(\phi^i, \partial_a \phi^i, x^a) & \rightarrow \mathcal{L}(\tilde{\phi}^i, \partial_a \tilde{\phi}^a, \tilde{x}^a), \label{L}\\
   \tilde{\phi}^i&=\phi^i - \delta\phi^i, \label{phi}\\
   \tilde{x}^a &= x^a-\delta x^a. \label{eq:x}
\end{align}
\end{subequations}
In order to find the generator of  transformations (\ref{phi}), (\ref{eq:x}), it is possible to consider the transformed prime derivatives of the field $\phi$, obtaining:
\begin{equation}\label{X1}
    X^{[1]} = \delta x^a \partial_a + \delta \phi^i \frac{\partial}{\partial \phi^i} + (\partial_a\delta\phi^i-\partial_a \phi^i\partial_b\delta x^b) \frac{\partial}{\partial(\partial_a\phi^i)},
\end{equation}
where the suffix $^{[1]}$ indicates the first prolongation of the Noether vector, including only the first derivatives of the fields. If the Euler-Lagrangian equations are invariant under transformations (\ref{phi}), (\ref{eq:x}), then there exists a function $h^a=h^a(x^a, \phi^i)$ such that:
\begin{equation}
    \tilde{\mathcal{L}}=\frac{d \tilde{x}^a}{dx^a} \mathcal{L} + \partial_a h^a.
\end{equation}
The two Lagrangians (i.e., the initial and the transformed ones) differ only by a four-divergence. Thus, substituting the first prolongation of the Noether vector (\ref{X1}), we get:
\begin{equation}\label{Noether}
    X^{[1]}\mathcal{L}+\partial_a\delta x^a \mathcal{L} = \partial_a h^a.
\end{equation}
In conclusion, if the Euler-Lagrangian equations are invariant under transformations (\ref{phi}), (\ref{eq:x}), then the condition (\ref{Noether}) is respected.

Therefore, the conserved current, associated to the symmetry transformation, results to be:
\begin{equation}
    j^a = \frac{\partial \mathcal{L}}{\partial (\partial_a\phi^i)} \delta\phi^i - \frac{\partial \mathcal{L}}{\partial (\partial_a \phi^i)}\partial_b\phi^i \delta x^b + \mathcal{L} \delta x^a - h^a,
\end{equation}
and the first Noether Theorem can be equivalently recast in terms of the first prolongation of the Noether vector: \emph{If the condition $X^{[1]}\mathcal{L}+\partial_a\delta x^a \mathcal{L} - \partial_a h^a=0$ holds, then the quantity $j^a$ is a first integral of the Equations of Motion (EoMs).} \cite{bajardi_capozziello_2022}.

By integrating $\partial_a j^a$ over the four-volume $\Omega$ and using the divergence theorem, one obtains the Noether charge:
\begin{equation}
    {\cal Q}= \int_\Omega \partial_a j^a d\Omega= \int_\Sigma j^a d\Sigma_a,
\end{equation}
which is the flow of $j^a$ and a scalar quantity under symmetry transformations.

The second Noether theorem relates the symmetries of a given action with the symmetries of the EoMs and it formally states: \emph{Let S be an action having an infinite-dimensional Lie algebra of infinitesimal symmetries linearly parameterized by m arbitrary functions, $f^k (x^a), k=0, 1, 2, ..$ and let $\Gamma^m_k \equiv \Gamma^m_k(x^a, \phi, \partial_a \phi)$  and  $\zeta^m_k \equiv \zeta^m_k(x^a, \phi, \partial_a \phi)$ be two generic functions, depending on the considered transformations. Then, there exists a set of functions $\Phi_m$ such that}:
\begin{equation}\label{IINoether}
    \Phi_m \Gamma^m_k = \partial_a {\Phi_m \zeta^m{}_k{}^a}.
\end{equation}
If the functions parameterizing the action are independent of the space-time coordinates, we can connect the first and the second Noether theorems as follows:
\begin{equation}\label{eq:con}
    \partial_a (\Phi_m \zeta^m_{}k{}^a \delta f^k - h^a)=0.
\end{equation}
If $\Phi_m$ are a set of $m$ Euler-Lagrange equations, then Eq. (\ref{eq:con}) is a conserved quantity; as consequence, the existence of a set of Euler-Lagrange equations implies the existence of a conserved quantity, without any further assumption. From this point of view, the first theorem provides the explicit form of the Noether current, and throughout the second theorem,  EoMs are related to such a current.

\subsection{Internal Symmetries}
\label{subsec:InternSy}

Let us consider now an on-shell internal symmetry transformation, so that space-time coordinates are not taken into account. Then, transformation laws (\ref{L}), (\ref{phi}), and (\ref{eq:x}) become:
\begin{subequations}
    \begin{align}
    \tilde{\mathcal{L}}(\tilde{x}^a, \tilde{\phi}^i, \partial_a\tilde{\phi}^i) &=\mathcal{L}(x^a, \partial_a\phi^i, \partial_a \phi^i),\\
    \delta \phi^i &= \tilde{\phi}^i - \phi^i,\\
    \delta x^a &=0,
    \end{align}
\end{subequations}
as well as the Lagrangian variation:
\begin{equation}
    \delta \mathcal{L}=\frac{\partial \mathcal{L}}{\partial \phi^i}\delta \phi^i + \frac{\partial \mathcal{L}}{\partial (\partial_a \phi^i)} \delta \partial_a \phi^i = \partial_a \Bigg(\frac{\partial \mathcal{L}}{\partial(\partial_a \phi^i)} \delta\phi^i \Bigg)=0.
\end{equation}
In this case, the Noether current and the Noether vector become:
\begin{subequations}
    \begin{align}
        j^a &=\frac{\partial \mathcal{L}}{\partial (\partial_a \phi^i)}\delta \phi^i, \label{eq:J} \\
        X &= \frac{\partial}{\partial \phi^i}\delta \phi^i + \frac{\partial}{\partial (\partial_a\phi^i)}\delta\partial_a\phi^i.
    \end{align}
\end{subequations}
Now, we can derive the cyclic variables of the system $\psi^1$, such that its conjugate momentum is a constant of motion:
\begin{equation}
    \frac{\partial \mathcal{L}}{\partial (\partial_a \psi^1)}= \pi^a_{\psi^1}.
\end{equation}
According to Eq. (\ref{eq:J}), $\pi^a_{\psi^1}$ is related to the Noether current, if the generator of $\psi^1$ is of the order of unity. Therefore, introducing a general change of variables $\phi^i \rightarrow \psi^i(\phi^j)$, such that $\psi^1$ is cyclic, we can rewrite the Noether vector $X$, obtaining the Noether current, which is expressed with respect to the fields $\psi^i$ as:
\begin{equation}\label{eq:curr}
    j^a=\delta\psi^i\frac{\partial\mathcal{L}}{\partial(\partial_a \psi^i)}=i_X d\psi^i \frac{\partial \mathcal{L}}{\partial(\partial_a \psi^i)}.
\end{equation}
The equality between the conserved quantity and the conjugate momentum of $\psi^1$, according to Eq (\ref{eq:curr}), implies that $\pi^a_{\psi^1}= cost$ (see Ref. \cite{bajardi_capozziello_2022}, for more details).

The condition $X' \mathcal{L'} = X \mathcal{L} = 0$ holds regardless of the variables considered, so that the Noether symmetry is preserved under the change of variables. If $T$ is a covariant tensor and $L_X T=0$, then we get the so-called Killing vector fields \cite{Manoff77, Woolley77}.

\subsection{Invariance conditions}
\label{subsec:invariace}

Let $X$ be a vector field on $M$ and $\omega$ a form on $M$. A differential form is conformal invariant with respect to a vector field $X$ if $L_X \omega = h \omega$, where $h\in \mathcal{C}^{\infty}(M)$. If $h=0$, then $L_X \omega=0$ and $\omega$ is invariant with respect to $X$, namely it does not change along the trajectories of $X$ \cite{Steeb78}. Therefore, $\omega$ is the related integral of $X$
\cite{Dieudonne681, Fels97, Steeb78}. 

The trajectories of the vector field $Y$ are invariant under a one-parameter group of transformations generated by a vector field $X$, if $L_X Y=[X, Y]=0$, representing an integrability condition.
Finally, by considering functions, i.e., 0-forms, a function $f$ is an invariant with respect to $X$ (first integral of the dynamical system $x = X(x)$), if $L_X f=0$. 
Thus, if $X$ is the generator of a certain symmetry, then the conserved quantities of the Noether theorem close on the same algebra as the generators:
%\begin{equation}
    %\begin{cases}
    \begin{eqnarray}
        \big[X^i, X^j \big]=i f^{ij}_k X^k 
        \\
        \{\Sigma^i, \Sigma^j \}= i f^{ij}_k\Sigma^k,
        \end{eqnarray}
    %\end{cases}
%\end{equation}
where the curly brackets are the Poisson brackets of the conserved quantities $\Sigma^i$. 

When the Lagrangian $\mathcal{L}$ is invariant under a transformation $\delta \mathcal{L}=0$, its internal symmetries are selected by the Noether vector $X$ and the related Noether current $j^a$ is a constant of motion (cf. Eq. (\ref{eq:J})).

The condition of internal symmetry $X\mathcal{L}=0$ is encoded into the Lie derivative, since:
\begin{equation}
    L_X \mathcal{L}=\big[X,\mathcal{L}\big]=X\mathcal{L}=0.
\end{equation}
Thus, the Lagrangian is invariant with respect to the vector field $X$, assuring the conserved quantity $\Sigma$, which  represents the Noether charge. Moreover, when external symmetries are considered, we can recast Eq. (\ref{Noether}) in terms of the first prolongation of the Noether vector, generator of the transformation, and the Lie derivative:
\begin{align}
    &X^{[1]}\mathcal{L}+\partial_a\delta x^a \mathcal{L} = \partial_a h^a \hookrightarrow \notag \\
    & L_X \mathcal{L} - \partial_b \delta x^b \mathcal{T} = \partial_a h^a, \label{eq:LieNoether}
\end{align}
where $\mathcal{T}$ is the trace of the energy-momentum tensor. By considering a time-dependent scalar field, Eq. (\ref{eq:LieNoether}) becomes:
\begin{equation}
    L_X \mathcal{L}+\dot{\xi}\mathcal{H}=\dot{h},
\end{equation}
with $\mathcal{H}$ the Hamiltonian of the system and $\xi=\delta t$. The choice of the function $h$ is a gauge choice, therefore if $\dot{\xi}=\xi_0$, then $L_X \mathcal{L} = c= costant$. If $c$ is trivial, i.e. $c=0$, internal symmetries are recovered.

In order to define the symmetries of the metric tensor $g$, we can apply the above results on it. A vector field $X$ on a (pseudo-)Riemannian manifold $(M,g)$ is a Killing field if
\begin{equation}
    L_{X} g = 0. 
\end{equation}
The Killing equation holds if and only if $g$ is invariant under the flow of $X$. More generally in fact:
\begin{equation}\label{eq:metric}
    L_X g_{\mu\nu}=2 g_{\mu\nu} \phi(x^\mu)
\end{equation}
Therefore, there are three types of Killing vectors, depending on the properties of $\phi$:
\begin{enumerate}
    \item Proper Killing Vector: $\phi(x^\mu)\neq 0 \rightarrow L_X g_{\mu\nu}=2g_{\mu\nu} \phi(x^\mu)$;
    \item Special Killing Vector: $\phi(x^\mu)=0 \rightarrow L_X g_{\mu\nu}=0$;
    \item Homotetic Killing Vector: $\partial_\mu \phi(x^\mu) \rightarrow L_X g_{\mu\nu} \sim g_{\mu\nu}$
\end{enumerate}
The special Killing vector encodes the concept of isometry on a given manifold.

Let $(M, g)$ be a (pseudo)-Riemannian manifold. A diffeomorphism $f: M \rightarrow M$ is an isometry if it preserves the metric tensor:
\begin{equation}\label{eq:isometry}
    f^{\ast} g_{f(p)} = g_{p},
\end{equation}
which implies $g_{f(p)} (f_{\ast}X, f_{\ast}Y) = g_p(X, Y)$ for $X, Y \in T_p M$.

If $\phi_t : M \rightarrow M$ is a one-parameter group of transformations, which generates the Killing vector field $X$, then, according to the definition of Killing vector, the local geometry does not change as we move along $\phi_t$. In this sense, the Killing vector fields represent the direction of the symmetry of a manifold \cite{Nakahara2003}. 

Finally, the Lie derivative can be used to find and select the symmetries of a given Lagrangian; in fact by considering Eq. (\ref{eq:LieNoether}), we can classify the Noether symmetries as \cite{bajardi_capozziello_2022}:
\begin{itemize}
    \item $L_X \mathcal{L} = \psi(x^a, \phi^i, \partial_a \phi^i)=\partial_b \delta x^b \mathcal{T} + \partial_a h^a$,   General Noether Symmetry;
    \item $L_X \mathcal{L}=cost$,  Noether Symmetry for canonical Lagrangians;
    \item $L_X \mathcal{L}=0$,  Internal Noether Symmetry.
\end{itemize}
Thus, the Killing vectors and conserved quantities obey the same rules imposed by the Lie derivative, applied to the metric tensor, and they can be used as criteria to determine the symmetry of a given dynamical system \cite{Camci2015, Grumiller14}.

\section{The Equivalence Principle as a Noether Symmetry}
\label{sec:EP and Noether Symmetry}
Previous considerations, in particular exploiting the second Noether theorem formulated with respect to the Lie derivative as in Eq. (\ref{eq:LieNoether}), can be specified in view to obtain the EP as a Noether symmetry.
Let us consider the Lagrangian $\mathcal{L}$ of a free particle
\begin{equation}
    \label{Lagrangian}
    \mathcal{L} = g_{\mu\nu} \Dot{x}^{\mu} \Dot{x}^{\nu} = g_{\mu\nu} u^{\mu} u^{\nu},
\end{equation}
where $u^{\mu}$ is a four-velocity. The Lie derivative of a function $f$ is defined as:
\begin{equation}
    L_{X}f = X \cdot \nabla f
\end{equation}
and represents the directional derivative of $f$ along $X$, namely the rate of change of $f$ measured by a comoving observer. By applying the Lie derivative to the Lagrangian (\ref{Lagrangian}), we have
\begin{equation}
    \label{Lie derivative L}
    L_{X}\mathcal{L}=  X \cdot \nabla \mathcal{L} = X^{\alpha} \nabla_{\alpha} (g_{\mu\nu} u^{\mu} u^{\nu}) = X^{\alpha} \nabla_{\alpha} (u^{\mu}u_{\mu}),
\end{equation}
which leads to
\begin{equation}
    \label{calcoli}
    L_{X}\mathcal{L} = X^{\alpha}\nabla_{\alpha}g_{\mu\nu} u^{\mu}u^{\nu} + 2X^{\alpha} u_{\mu} \nabla_{\alpha}u^{\mu}. 
\end{equation}
When we work in a metric affine theory, we consider the most general linear affine connection, defined in Eq. (\ref{affine connection}), with its related dynamical variable of non-metricity, given by  Eq. (\ref{eq:nonmetricity}).
Therefore, when the non-metricity tensor is considered, we can rewrite Eq. (\ref{calcoli}) as:
\begin{equation} \label{nm}
    L_{X}\mathcal{L} = X^{\alpha} Q_{\alpha \mu \nu} u^{\mu}u^{\nu} + 2X^{\alpha} u_{\mu} \nabla_{\alpha}u^{\mu}
\end{equation}
The covariant derivative of the metric tensor can be expressed as follows:
\begin{align}\label{Q}
    Q_{\alpha \mu\nu} = & \partial_{\alpha} g_{\mu\nu} - \Gamma^{\lambda}_{\alpha\mu}g_{\lambda \nu} - \Gamma^{\lambda}_{\alpha \nu} g_{\lambda\mu}\notag = \\ & \partial_{\alpha} g_{\mu\nu} - \tilde{\Gamma}^{\lambda}_{\alpha\mu}g_{\lambda \nu} - K^{\lambda}_{\alpha\mu}g_{\lambda \nu} - L^{\lambda}_{\alpha\mu}g_{\lambda \nu}\notag - \\ & \tilde{\Gamma}^{\lambda}_{\alpha \nu} g_{\lambda\mu} - K^{\lambda}_{\alpha \nu} g_{\lambda\mu} - L^{\lambda}_{\alpha \nu} g_{\lambda\mu}\notag = \\ - &  K^{\lambda}_{\alpha\mu}g_{\lambda \nu} - L^{\lambda}_{\alpha\mu}g_{\lambda \nu} - K^{\lambda}_{\alpha \nu} g_{\lambda\mu} - L^{\lambda}_{\alpha \nu} g_{\lambda\mu}
\end{align}
Eq. (\ref{Q}) can be inserted into Eq. (\ref{nm}) obtaining:
\begin{equation}
    \label{Lie derivative generale}
   L_{X}\mathcal{L} = 2X^{\alpha} u_{\mu} \nabla_{\alpha} u^{\mu} - 2X^{\alpha} (K^{\lambda}_{\alpha\mu} + L^{\lambda}_{\alpha\mu})u^{\mu} u_{\lambda}.
\end{equation}
According to Eq. (\ref{Lie derivative generale}), it is possible to reconstruct the geodesic, that is the autoparallel equations. In particular, since the four-velocity is the tangent vector to the worldline $x^{\mu}$, and $u_{\mu} = dx_{\mu}/d\tau$, with $\tau$ the proper time, the Lie vector $X^{\mu}$ exactly corresponds to the four-velocity $u^{\mu}$. Thus, we have:
\begin{align}
    L_{X}\mathcal{L} &=  2u^{\alpha} u_{\mu} \nabla_{\alpha} u^{\mu} - 2u^{\alpha} (K^{\lambda}_{\alpha\mu} + L^{\lambda}_{\alpha\mu})u^{\mu} u_{\lambda} \rightarrow \label{eq:KL} \\ 
    L_{X}\mathcal{L} &= 2u^{\alpha} u_{\mu} \nabla_{\alpha} u^{\mu} - 2u^{\alpha}(\Gamma^{\lambda}_{\alpha\mu} - \tilde{\Gamma}^{\lambda}_{\alpha\nu})u^{\mu}u_{\lambda} \label{eq:gamma},
\end{align}
the Lie derivative of a free particle Lagrangian for a general affine connection represents how much the theory is different from GR.
Moreover, we can define:
\begin{subequations}
\begin{align} 
a^\mu&:=u^\lambda\nabla_\lambda u^\mu,\label{eq:acc}\\
\tilde{a}_\mu&:=u^\lambda \nabla_\lambda u_\mu=a_\mu+Q_{\lambda\nu\mu}u^\lambda u^\nu,\label{eq:an_acc}
\end{align}
\end{subequations}
where $a^\mu$ is an \emph{acceleration}, whereas $\tilde{a}_\mu$ is an \emph{anomalous acceleration}. We  recover the STG results \cite{Ferrara22} since:
\begin{align}\label{non metricity LD}
    L_{X}\mathcal{L} = &Q_{\alpha\mu\nu} u^{\alpha}u^{\mu}u^{\nu} + 2u^{\alpha}u_{\mu}\nabla_{\alpha}u^{\mu}\notag =  \\
    &Q_{\alpha\mu\nu} u^{\alpha}u^{\mu}u^{\nu} + 2 u_{\mu}a^{\mu} = a^{\mu}u_{\mu} + \tilde{a}_{\mu}u^{\mu}
\end{align}
The non-metricity tensor expresses how much the anomalous acceleration deviates from the standard acceleration, and it is also responsible of how much the acceleration departs from the spatial hypersurface orthogonal to the four-velocity. Moreover, the Lie derivative of the free-particle Lagrangian encodes the feature that the non-metricity does not preserve the norm and the length of a vector and, as consequence, its behaviour during the parallel transport.
Furthermore, developing the covariant derivative of $a^{\nu}$, according to Eq. (\ref{eq:acc}), we also have: 
\begin{equation}\label{derivata parziale a}
    u^{\alpha}\nabla_{\alpha}u^{\nu} = u^{\alpha} \Big[ \frac{\partial u^{\nu}}{\partial x^{\alpha}} + \Gamma^{\nu}_{\alpha\lambda}u^{\lambda} \Big]
\end{equation}
and inserting Eq. (\ref{derivata parziale a}) into Eq. (\ref{non metricity LD}), we obtain:
\begin{equation}
Q_{\alpha\mu\nu}u^{\alpha}u^{\mu}u^{\nu} + 2 \Big[\frac{d^{2}x^{\nu}}{d\tau^{2}} + \Gamma^{\nu}_{\alpha\lambda}u^{\alpha}u^{\lambda} \Big] u_{\nu} = a^{\nu}u_{\nu} + \tilde{a}_{\nu}u^{\nu}    
\end{equation}
If we develop non-metricity tensor according to Eq. (\ref{Q}), we obtain:
\begin{align}\label{geo}
    &Q_{\alpha\mu\nu}u^{\alpha}u^{\mu}u^{\nu} + 2 \Big[\frac{d^{2}x^{\nu}}{d\tau^{2}} + \Gamma^{\nu}_{\alpha\lambda}u^{\alpha}u^{\lambda} \Big] u_{\nu}  =\notag\\ 
    &(\partial_{\alpha}g_{\mu\nu} - \Gamma^{\lambda}_{\alpha\mu}g_{\lambda\nu} - \Gamma^{\lambda}_{\alpha\nu} g_{\mu\lambda}) u^{\alpha}u^{\mu}u^{\nu} + \\ 
    &2 \Big[\frac{d^{2}x^{\nu}}{d\tau^{2}} + \Gamma^{\nu}_{\alpha\lambda}u^{\alpha}u^{\lambda} \Big] u_{\nu} =\notag\\
    &\frac{d^{2}x^{\nu}}{d\tau^{2}}u_{\nu} + \tilde{\Gamma}^{\nu}_{\alpha\mu}u^{\alpha}u^{\mu}u_{\nu}=\frac{1}{2} \Big( a^{\nu}u_{\nu} + \tilde{a}_{\nu}u^{\nu} \Big)
\end{align}
Thus, we have the autoparallel equation of STEGR for a trivial acceleration $a^\mu=0$, that is Eq. (\ref{eq:acc}):
\begin{equation}
    L_{X}\mathcal{L} = Q_{\alpha\mu\nu} u^{\alpha}u^{\mu}u^{\nu} = \tilde{a}_{\nu}u^{\nu}.
\end{equation}
As consistency check, we rewrite Eq. (\ref{geo}) as
\begin{align}
    &\frac{d^{2}x^{\nu}}{d\tau^{2}}u_{\nu} + \tilde{\Gamma}^{\nu}_{\alpha\mu}u^{\alpha}u^{\mu}u_{\nu} = \frac{1}{2}\tilde{a}_{\nu}u^{\nu} \\
    &\frac{d^{2}x^{\nu}}{d\tau^{2}}u_{\nu} + \tilde{\Gamma}^{\nu}_{\alpha\mu}u^{\alpha}u^{\mu}u_{\nu} = \frac{1}{2} Q_{\alpha\mu\nu}u^{\alpha}u^{\mu}u^{\nu} \\
    &\frac{d^{2}x^{\nu}}{d\tau^{2}}u_{\nu} + \tilde{\Gamma}^{\nu}_{\alpha\mu}u^{\alpha}u^{\mu}u_{\nu} + [K^{\nu}_{\alpha\mu}+L^{\nu}_{\alpha\mu}]u^{\alpha}u^{\mu}u_{\nu} =0 
\end{align}
Therefore, in the general case, the Lie derivative of the  Lagrangian $\mathcal{L}$ is equal to the autoparallel equation of a  metric affine theory:
\begin{equation}\label{eq:MAG parallel}
    L_{X}\mathcal{L} = \frac{d^{2} x^{\nu}}{d\tau^{2}} + \Gamma^{\nu}_{\alpha\mu} u^{\alpha}u^{\mu}=0
\end{equation}
 GR is recovered if we have that Eq. (\ref{nm}) implies the condition:
\begin{subequations}
\begin{align}
    L_{X} \mathcal{L} &= 0   \label{EP}  \\
    Q_{\alpha \mu \nu} = 0 & \iff X^{\alpha}\nabla_{\alpha}u^{\mu} = 0, \label{EP metric}
\end{align}
\end{subequations}
since the covariant derivative along the direction of the vector $X$ is zero and we again recover the geodesic transport equation:
\begin{align} \label{lie}
    &X^{\alpha}\nabla_{\alpha}u^{\mu} \equiv\notag  \\ 
    &\frac{dx^{\alpha}}{d\lambda} \Big[\frac{\partial u^{\mu}}{\partial X^{\alpha}} + \tilde{\Gamma}^{\mu}_{\alpha \rho} u^{\rho} \Big] = \frac{du^{\mu}}{d\lambda} + \tilde{\Gamma}^{\mu}_{\alpha \rho} u^{\rho} X^{\alpha}.
\end{align}
In other words, we restore  the norm conservation of a vector and the geodesic of GR:
\begin{equation}
    \frac{d^2 x^{\mu}}{d \tau^{2}} + \tilde{\Gamma}^{\mu}_{\rho\nu} \dot{x}^{\rho} \dot{x}^{\nu} = 0
\end{equation}
Thus, the EEP (and SEP) is derived by the Noether symmetry of  Lagrangian \eqref{Lagrangian}. On the contrary, the SEP (and  EEP) violation is related to  the non-metricity tensor thanks to  Eqs. (\ref{eq:KL}) and (\ref{eq:gamma}). For a non-metric theory, the only way to recover EEP is to choose the coincident gauge, nullifying the general affine connection. In this case, Eq. (\ref{eq:MAG parallel}) reduces to:
\begin{equation}
    L_{X}\mathcal{L} = \frac{d^{2} x^{\nu}}{d\tau^{2}} =0\,.
\end{equation}
We can conclude that the SEP is a Noether symmetry since we locally recovered the SR laws. This is a necessary and sufficient condition. As a consequence also  EEP is a Noether symmetry and, also in this case,  it is a necessary and sufficient condition.

It is worth noticing that the above two accelerations  can be recovered also considering  the Raychaudhuri equation in spacetimes with torsion and non-metricity as reported in Ref.\cite{Iosifidis}. In that case, they are  called  path (because  related to autoparallels) and hyper-acceleration
respectively. The latter corresponds to our  anomalous acceleration.

\section{Discussion and Conclusions}
\label{Conc}

In this work, we showed that the EEP is a Noether Symmetry for metric-affine theories of gravity. In particular, this statement holds for  GR. 

We introduced  MAG as  possible extensions of GR. In this general picture,  dynamics  can be related to  torsion  $T^{\alpha}_{\mu\nu}$ and non-metricity  $Q_{\alpha\mu\nu}$, besides  curvature $R^{\alpha}_{\beta\mu\nu}$. In particular, we focused on the dynamical equivalence of GR, TEGR, and STEGR and the role acquired by  EEP  in the geometric trinity of gravity. It is recovered both in TEGR and STEGR through a gauge choice of the affine connection. Moreover,     symmetric-teleparallel theories of gravity, more general than STEGR,  can be characterized  by a non-trivial non-metricity tensor.

Considering  connection and parallel transport,   EEP is strictly related to the Principle of General Covariance and the presence of autoparallel curves. These geometric concepts can be reinterpreted thanks to the two Noether theorems which allow  to find conserved quantities related to  internal and space-time symmetries. 

In this framework, it is possible to demonstrate that the  EP (in particular the EEP and SEP) is a Noether symmetry in agreement with the formulation  with respect to the Fermi coordinates (Fermi-Walker transport). Specifically, considering local coordinates, we can take into account the Lagrangian formalism and  follow the trajectory of  particles along  world lines.  We obtain that, in a general metric-affine context, the parallel transport does not preserve orthonormal vectors so that one cannot ask for  the conservation of metric along  timelike geodesics. However, it is possible to recover EEP when the coincident gauge is chosen. 

The validity of SEP implies the validity of EEP. In fact SEP is an extensions of EEP including  gravitational phenomena. In absence of gravity, the physics reduces to SR, giving a local Minkowskian space-time structure. According to EEP, a general affine connection $\Gamma$ is locally indistinguishable from the flat Levi–Civita connection $\tilde{\Gamma}$ of the Minkowski metric $\eta_{\alpha\beta}$. In GR case, this means that the  $\Gamma$ itself is the Levi-Civita connection. Here, we demonstrated that this approach can be developed also for  the symmetric teleparallel theories through the  coincident gauge \cite{Hohmann21}. It is worth noticing that such a gauge allows also to recover gravitational waves strating from general non-metric theories of gravity \cite{Capozziello2024vix}.

In this perspective,  several fundamental issues can be  investigated. For example, since SEP can be recovered in TEGR and STEGR, any possible violation (e.g. at quantum level) \cite{Altschul14, Tino20}, would make TEGR and STEGR more fundamental  than GR, because they do not require EP as a basic principle. Moreover,  it is essential to establish  the true number of degrees of freedom of gravitational field. This issue  is  still matter of debate  and should be fully investigated by precision experiments \cite{DiVirgilio21}, gravitational wave astronomy \cite{Abedi17}, and precision cosmology observations \cite{Cai15, Abdalla22, Mehdi}.

Furthermore,  the equivalence among different theories  could be restored by considering appropriate boundary terms, which, again, are fixed by  gauge choices. In particular, $f(R)$, $f(T,B)$, and $f(Q,B)$ can be compared as fourth order theories when  appropriate boundary terms $B$ are defined in each gravity framework \cite{Bahamonde15,Bahamonde17,Bahamonde22,Capozziello18,DeFalco23}. Thus, SEP and the boundary terms $B$ could be  essential to establish the  correct relation among gravity theories. 

Finally, since  possible violations of the EEP (and SEP) are related to  non-metricity,  the experimental  investigation  of non-metricity tensor, and its related quantities, could be the key ingredient to establish the validity of EP both at classical and quantum level.

\section*{Acknowledgements}
The Authors  thank Vittorio De Falco for the useful comments and discussions on the topics and  acknowledge the {\it Istituto Italiano di Fisica Nucleare} (INFN) {\it iniziative specifiche} QGSKY and MOONLIGHT2.
This paper is based upon work from COST Action CA21136 {\it Addressing observational tensions in cosmology with systematics and fundamental physics} (CosmoVerse) supported by COST (European Cooperation in Science and Technology). 
\medskip
\bibliography{biblo}

\end{document}